\documentclass[a4paper]{article}
\usepackage{RR}
\RRNo{5972}
\usepackage{hyperref}

\usepackage{t1enc} 
\usepackage{amsmath}
\usepackage{latexsym}
\usepackage{theorem}
\usepackage{amssymb}
\usepackage{url}

\newcommand{\comment}[1]{}

\newcommand{\vsp}[1][3mm]{\vspace*{#1}}
\newcommand{\moins}{\setminus}
\newcommand{\vide}{\emptyset}

\newcommand{\eg}{{\em e.g.} }

\newcommand{\xu}{\{x\to u\}}
\newcommand{\xv}{\{x\to v\}}
\newcommand{\xw}{\{x\to w\}}

\newcommand{\dom}{\mr{dom}}
\newcommand{\codom}{\mr{codom}}
\newcommand{\FV}{\mr{FV}}
\newcommand{\pos}{\mr{Pos}}

\renewcommand{\a}{\rightarrow}
\newcommand{\A}{\Rightarrow}

\renewcommand{\to}{\mapsto}

\newcommand{\I}[1]{[\![#1]\!]}

\newcommand{\ex}{\exists}
\newcommand{\all}{\forall}
\newcommand{\ou}{\vee}

\newcommand{\et}{\wedge}

\newcommand{\sle}{\subseteq}
\newcommand{\sge}{\supseteq}

\newcommand{\tle}{\unlhd}
\newcommand{\tge}{\unrhd} 
\newcommand{\tlt}{\lhd}
\newcommand{\tgt}{\rhd}

\newcommand{\lex}{_\mr{lex}}
\newcommand{\mul}{_\mr{mul}}

\renewcommand{\o}[1]{{\overline{#1}}}

\renewcommand{\b}{\beta}

\renewcommand{\d}{\delta}

\newcommand{\vep}{\varepsilon}

\renewcommand{\t}{\theta}

\renewcommand{\l}{\lambda}

\newcommand{\s}{\sigma}
\renewcommand{\S}{\Sigma}

\newcommand{\w}{\omega}

\newcommand{\mi}{\mathit}
\newcommand{\mc}{\mathcal}

\newcommand{\mr}{\mathrm}
\newcommand{\mb}{\mathbb}

\newcommand{\mf}{\mathfrak}
\newcommand{\ms}{\mathsf}

\newcommand{\bN}{\mb{N}}

\newcommand{\bT}{\mb{T}}

\newcommand{\cA}{\mc{A}}
\newcommand{\cB}{\mc{B}}
\newcommand{\cC}{\mc{C}}
\newcommand{\cD}{\mc{D}}

\newcommand{\cF}{\mc{F}}

\newcommand{\cI}{\mc{I}}

\newcommand{\cQ}{\mc{Q}}

\newcommand{\cT}{\mc{T}}

\newcommand{\cV}{\mc{V}}

\newcommand{\cX}{\mc{X}}

\newcommand{\ka}{\mf{a}} 

\newcommand{\fB}{\ms{B}}

\newcommand{\fD}{\ms{D}}

\newcommand{\fL}{\ms{L}}

\newcommand{\fP}{\ms{P}}

\newcommand{\fR}{\ms{R}}

\newcommand{\va}{{\vec{a}}}
\newcommand{\vb}{{\vec{b}}}

\newcommand{\vl}{{\vec{l}}}

\newcommand{\vt}{{\vec{t}}}
\newcommand{\vu}{{\vec{u}}}
\newcommand{\vv}{{\vec{v}}}
\newcommand{\vw}{{\vec{w}}}

\newcommand{\vA}{{\vec{A}}}
\newcommand{\vB}{{\vec{B}}}

\newcommand{\vT}{{\vec{T}}}
\newcommand{\vU}{{\vec{U}}}
\newcommand{\vV}{{\vec{V}}}

\newenvironment{rul}
  {$\begin{array}{rcl}}
  {\end{array}$}

\newenvironment{rew}[1][~~\a~~]
  {$\begin{array}{r@{#1}l}}
  {\end{array}$}
\newenvironment{rewc}[1][~~\a~~]
  {\begin{center}\begin{rew}[#1]}
  {\end{rew}\end{center}}

  {\end{array}$\end{center}}

\newcounter{counter}

\newcounter{explnum}
{\theorembodyfont{\rmfamily} 
  \newtheorem{definition}[counter]{Definition}
  \newtheorem{lemma}[counter]{Lemma}
  \newtheorem{theorem}[counter]{Theorem}

  \newtheorem{example}[explnum]{Example}
}

\newcommand{\qed}{\hfill$\blacksquare$} 
\newenvironment{proof}{{\bf Proof.}}{}

\newenvironment{lstgeneric}[2]
  {\begin{list}{#1}{\topsep=.5mm\itemsep=.5mm\parsep=0mm%
    \itemindent=-3ex\labelsep=1ex\labelwidth=0ex #2}}
  {\end{list}}

\newenvironment{lst}[1]
  {\begin{lstgeneric}{#1}{\itemindent=-1ex}}
  {\end{lstgeneric}}

\newenvironment{enumi}[1]
  {\begin{lstgeneric}{}{\usecounter{enumi}\leftmargin=7mm%
    }}
  {\end{lstgeneric}}

\newenvironment{enumalphai}
  {\begin{lstgeneric}{}{\usecounter{enumi}\leftmargin=7mm%
    }}
  {\end{lstgeneric}}

\newcommand{\SN}{\mr{SN}}
\newcommand{\CC}{\mr{CC}}
\newcommand{\CR}{\mr{CR}}
\newcommand{\Acc}{\mr{Acc}}

\newcommand{\rpo}{_{\mr{rpo}}}
\newcommand{\rco}{_{\mr{rco}}}
\newcommand{\horpo}{_{\mr{horpo}}}
\newcommand{\horco}{_{\mr{horco}}}
\newcommand{\whorco}{_{\mr{whorco}}}
\newcommand{\stat}[1]{_{\mr{stat}_{#1}}}

\newcommand{\red}[1]{{\a\!\!(#1)}}

\newcommand{\ab}{\a_\b}

\newcommand{\lx}{\l x}
\newcommand{\ly}{\l y}

\RRdate{30 August 2006}
\RRauthor{
Fr\'ed\'eric Blanqui
}
\authorhead{Fr\'ed\'eric Blanqui}
\RRtitle{(HO)RPO Revisit\'e}
\RRetitle{(HO)RPO Revisited}
\titlehead{(HO)RPO Revisited}
\RRresume{La notion de cl\^oture de calculabilit\'e a \'et\'e introduite
pour prouver la terminaison de la combinaison de r\'ecriture d'ordre
sup\'erieur et de beta-r\'eduction. Elle est aussi utilis\'ee pour
enrichir l'ordre r\'ecursif sur les chemins (RPO) \`a l'ordre
sup\'erieur (HORPO). Dans cet article, nous \'etudions la relation
entre la cl\^oture de calculabilit\'e et (HO)RPO. Nous montrons que
RPO est \'egal \`a un ordre naturellement d\'efinit \`a partir de la
cl\^oture. A l'ordre sup\'erieur, nous obtenons un ordre contenant
HORPO dont la preuve de bonne fondation repose sur la correction de la
cl\^oture. Cela fournit une mani\`ere simple d'\'etendre HORPO \`a des
syst\`emes de types plus riches.}
\RRabstract{The notion of computability closure has been introduced
for proving
the termination of the combination of higher-order rewriting and
beta-reduction. It is also used for strengthening the higher-order
recursive path ordering. In the present paper, we study in more
details the relations between the computability closure and the
(higher-order) recursive path ordering. We show that the first-order
recursive path ordering is equal to an ordering naturally defined from
the computability closure. In the higher-order case, we get an
ordering containing the higher-order recursive path ordering whose
well-foundedness relies on the correctness of the computability
closure. This provides a simple way to extend the higher-order
recursive path ordering to richer type systems.}
\RRmotcle{terminaison, ordre, lambda-calcul, r\'e\'ecriture}
\RRkeyword{termination, ordering, lambda-calculus, rewriting}
\RRprojet{Protheo}  
\RRtheme{\THSym}
\URLorraine 
\begin{document}
\makeRR   


\section{Introduction}

We are interested in automatically proving the termination of the
combination of $\b$-reduction and higher-order rewrite rules. There
are two important approaches to higher-order rewriting: rewriting on
$\b\eta$-equivalence classes (or $\b\o\eta$-normal forms)
\cite{mayr98tcs} with higher-order pattern-matching (higher-order
unification on higher-order patterns has been proved decidable in
\cite{miller89elp}), and the combination of $\b$-reduction and term
rewriting with higher-order pattern-matching \cite{klop93tcs}. The
relation between both has been studied in \cite{oostrom93hoa}. The
second approach is more atomic since a rewrite step in the first
approach can be directly encoded by a rewrite step together with
$\b$-steps in the second approach. In this paper, we consider the
second approach, restricted to first-order pattern-matching (we do not
have abstractions in rule left-hand side).

The combination of $\b$-reduction and rewriting is naturally used in
dependent type systems and proof assistants implementing the
proposition-as-type and proof-as-object paradigm. In these systems,
two propositions equivalent modulo $\b$-reduction and rewriting are
considered as equivalent (\eg $P(2+2)$ and $P(4)$). This is essential
for enabling users to formalize large proofs with many computations,
as recently shown by Gonthier and Werner's proof of the Four Color
Theorem in the Coq proof assistant. However, for the system to be able
to check the correctness of user proofs, it must at least be able to
check the equivalence of two terms. Hence, the necessity to have
termination criteria for the combination of $\b$-reduction with a set
$R$ of higher-order rewrite rules.

To our knowledge, the first termination criterion for such a
combination is Jouannaud and Okada's General Schema
\cite{jouannaud91lics,jouannaud97tcs}. It is based on Tait's technique
for proving the strong normalization of the simply-typed
$\l$-calculus \cite{tait72lc}. Roughly speaking, since proving the
strong $\b$-normalization of simply-typed $\l$-terms by induction on
the term structure does not work directly, Tait's idea was to prove a
stronger property that he called {\em strong computability}. Extending
Tait's technique to higher-order rewriting consists in proving that
function symbols are computable too, that is, that every function call
is computable whenever its arguments so are. This naturally leads to
the following question: which operations preserve computability? From
a set of such operations, one can define the {\em computability
closure} of a term $t$, written $\CC_R(t)$, as the set of terms that are
computable whenever $t$ so is. Then, to get normalization, it suffices
to check that, for every rule $f\vl\a r$, $r$ belongs to the
computability closure of $\vl$. The General Schema was implicitly
doing this. The first definition of computability closure appeared in
an 1997 unpublished note of Jouannaud and Okada which served as a
basis for \cite{blanqui99rta}, an extension to dependent types of the
computability closure. The computability closure was later extended to
higher-order pattern-matching \cite{blanqui00rta}, type-level
rewriting \cite{blanqui01lics,blanqui05mscs} and rewriting modulo AC
\cite{blanqui03rta}. Examples of computability-preserving operations are:

\begin{lst}{--}
\item application: if $u\in\CC_R(t)$ and $v\in\CC_R(t)$, then $uv\in\CC_R(t)$).
\item abstraction: if $u\in\CC_R(t)$, then $\lx u\in\CC_R(t)$).
\item recursive calls on structurally smaller arguments:
if $\vu\in\CC_R(f\vt)$ and $\vu\tlt\vt$, then $f\vu\in\CC_R(f\vt)$.
\item reduction: if $u\in\CC_R(t)$ and $u\a_R v$, then $v\in\CC_R(t)$.
\end{lst}

Another way to prove the termination of a set of rules is to find a
decidable well-founded rewrite relation containing these rules. A well
known such relation in the first-order case is the (inductively
defined) recursive path ordering \cite{plaisted78tr,dershowitz82tcs}
whose well-foundedness proof was initially based on Kruskal theorem
\cite{kruskal60ams}. The first attempts
\cite{loria92ctrs,lysne95rta,jouannaud96rta} made for generalizing
this ordering to the higher-order case were not able to orient G\"odel
system T for instance. Finally, in 1999, Jouannaud and Rubio succeeded
in defining such an ordering \cite{jouannaud99lics} by following the
termination proof technique developed in \cite{jouannaud97tcs}. By
the way, this provided the first well-foundedness proof of RPO not
based on Kruskal theorem. HORPO has also been extended to dependent
types later in \cite{walukiewicz03jfp}.

Although the computability closure on one hand, and the recursive path
ordering on the other hand, shares the same computability-based
techniques, there has been no precise comparison between these two
termination criteria. In \cite{walukiewicz03thesis}, one can find
examples of rules that are accepted by one criterion but not the
other. And Jouannaud and Rubio themselves use the notion of
computability closure for strengthening their ordering.

In the present paper, we explore the relations between both
criteria. We start from the trivial remark that the computability
closure itself defines an ordering: $t>_R u$ if $t=f\vt$ and
$u\in\CC_R(\vt)$. Proving the well-foundedness of this ordering simply
consists in proving that the computability closure is correct. Then,
we remark that $>_R$ is monotone and continuous for inclusion wrt
$R$. Thus, the computability closure admits a fixpoint which is a
well-founded ordering. In the first case order, we prove that this
ordering is the recursive path ordering. In the higher-order case, we
prove that we get an ordering containing HORPO. Although, we do not
get in this case a better definition, it shows that the
well-foundedness of HORPO can be reduced to the correctness of the
computability closure. This also provide a way to easily strengthen
HORPO. Another advantage of this approach is that it can easily be
extended to more complex type systems.


\section{First-order case}

To illustrate our approach, we first begin by presenting the
first-order case which is interesting on its own.

We assume given a set $\cX$ of variables and a disjoint set $\cF$ of
function symbols. Let $\cT$ be the set of first-order algebraic terms
built from $\cF$ and $\cX$ as usual. Let $\cV(t)$ (resp. $\cF(t)$) be
the set of variables (resp. symbols) occurring in $t$.

We assume given a {\em precedence} $\ge_\cF$ on $\cF$, that is, a
quasi-ordering whose strict part ${>_\cF}={\ge_\cF\moins\le_\cF}$ is
well-founded. Let ${\simeq_\cF}={\ge_\cF\cap\le_\cF}$ be its
associated equivalence relation.

A precedence can be seen as a particular case of quasi-ordering on
terms looking at top symbols only. We could extend our results to this
more general case, leading to extensions of the semantic path
ordering. See \cite{kamin80} for the first-order case, and
\cite{borralleras01lpar} for the higher-order case.

We assume that every symbol $f\in\cF$ is equipped with a {\em status}
$\mr{stat}_f\in \{\mr{lex},\mr{mul}\}$ defining how the arguments of
$f$ must be compared: lexicographically (from left to right, or from
right to left) or by multiset. We also assume that
$\mr{stat}_f=\mr{stat}_g$ whenever $f\simeq_\cF g$.

\begin{definition}
Given a relation $>$ on terms, let $(f,\vt)>\stat{} (g,\vu)$ iff
either $f>_\cF g$ or $f\simeq_\cF g$ and $\vt>\stat{f}^+\vu$.
\end{definition}

The ordering $>\stat{}$ is well-founded whenever $>$ so is ($>_\cF$ is
well-founded).

As usual, the set $\pos(t)$ of {\em positions} in a term $t$ is
defined as words on positive integers. If $p\in\pos(t)$, then $t|_p$
is the {\em subterm} of $t$ at position $p$, and $t[u]_p$ is the term
$t$ with $t|_p$ replaced by $u$. Let $\tle$ be the subterm relation.

A relation $>$ on terms is {\em stable by substitution} if $t\t>u\t$
whenever $t>u$. It is {\em stable by context} if $C[t]_p>C[u]_p$
whenever $t>u$. It is a {\em rewrite relation} if it is both stable by
substitution and context. Given a relation on terms $R$, let $\a_R$ be
the smallest rewrite relation containing $R$, $R^+$ be the transitive
closure of $R$, and $\SN(R)$ be the set of terms that are strongly
normalizing for $R$.


\begin{figure}[ht]
\centering\caption{First-order computability closure\label{fig-focc}}\vsp
\fbox{\begin{minipage}{10cm}
\centering\vsp
(arg)\quad $t_i\in \CC_R^f(\vt)$\\[2mm]

(decomp)\quad $\cfrac{g\vu\in \CC_R^f(\vt)}{u_i\in \CC_R^f(\vt)}$\\[2mm]

(prec)\quad $\cfrac{f>_\cF g\quad \vu\in \CC_R^f(\vt)}
{g\vu\in \CC_R^f(\vt)}$\\[2mm]

(call)\quad $\cfrac{f\simeq_\cF g\quad
\vu\in \CC_R^f(\vt)\quad \vt~(\a_R^+\cup\,\tgt)\stat{f}~\vu}
{g\vu\in \CC_R^f(\vt)}$\\[2mm]

(red)\quad $\cfrac{u\in \CC_R^f(\vt)\quad u\a_R^+ v}{v\in \CC_R^f(\vt)}$\vsp
\end{minipage}}
\end{figure}


Hereafter is a definition of computability closure similar to the one
given in \cite{blanqui99rta} except that:

\begin{lst}{--}
\item it is restricted to untyped first-order terms,
\item we abstracted away the set $R$ of rules and explicitly put it as argument
of the computability closure,
\item we added $\a_R^+$ for comparing arguments in (call).
\end{lst}

The main novelty is the addition of $\a_R^+$ in (call). This allows us
to get the recursive behavior of RPO: one can use the ordering itself
for comparing the arguments of a recursive call. The fact that this is
a computability-preserving operation was implicit in
\cite{blanqui99rta}. A complete proof of this fact for the higher-order
case is given in Lemma \ref{lem-hocr-wf}.

\begin{definition}[Computability closure]
Let $R$ be a relation on terms. The {\em computability closure} of a
term $f\vt$, written $\CC_R^f(\vt)$, is inductively defined in Figure
\ref{fig-focc}. Let $\CR(R)$ be the set of pairs $(f\vt,u)$ such that
$u\in\CC_R^f(\vt)$.
\end{definition}

One can easily prove that $\CR$ is monotone and $\w$-sup-continuous
for inclusion. It has therefore a least fixpoint that is reachable by
iteration from $\vide$.

\begin{definition}[Computability ordering]
Let the {\em first-order recursive computability ordering} $>\rco$
be the least fixpoint of $\CR$.
\end{definition}

Note that one gets the same ordering by replacing in (red) $\a_R^+$ by
$R$, and in (call) $\a_R^+\cup\,\tgt$ by $R$.


\begin{lemma}
\label{lem-foco-prop}
$>\rco$ is a transitive rewrite relation containing subterm.
\end{lemma}

\begin{proof}
Since $\CR$ is $\w$-sup-continuous and preserves the stability by
substitution, $>\rco$ is stable by substitution. For the transitivity,
assume that $t>\rco u>\rco v$. Then, $t$ must be of the form $f\vt$ and,
by (red), $t>\rco v$. For the stability by context, let $v=f\va t\vb$
and $t>\rco u$. By (arg), $v>\rco \va t\vb$. By (red), $v>\rco u$. Thus,
$\va t\vb ~(>\rco)\stat{f}~ \va u\vb$ and, by (call), $v>\rco f\va
u\vb$. Finally, $>\rco$ contains subterm by (arg).\qed
\end{proof}

It follows that (decomp) is derivable from (arg) and transitivity. We
introduce in Figure \ref{fig-foco} an inductive formulation of $>\rco$
obtained by replacing in the rules defining the computability closure
$u\in\CC_R^f(\vt)$ by $f\vt>\rco u$, and $R$ by $>\rco$.

This simple change in notations clearly shows that $\rco$ is equal to
$>\rpo$, whose definition is recalled in Figure \ref{fig-rpo}.


\begin{figure}[ht]
\centering\caption{First-order recursive computability ordering\label{fig-foco}}\vsp
\fbox{\begin{minipage}{10cm}
\centering\vsp
(arg)\quad $f\vt>\rco t_i$\\[2mm]


(prec)\quad $\cfrac{f>_\cF g\quad f\vt>\rco \vu}{f\vt>\rco g\vu}$\\[2mm]

(call)\quad $\cfrac{f\simeq_\cF g\quad f\vt>\rco \vu\quad
\vt~(>\rco)\stat{f}~\vu}{f\vt>\rco g\vu}$\\[2mm]

(red)\quad $\cfrac{f\vt>\rco u\quad u>\rco v}{f\vt>\rco v}$\vsp
\end{minipage}}
\end{figure}


\begin{figure}[ht]
\centering\caption{First-order recursive path ordering\label{fig-rpo}}\vsp
\fbox{\begin{minipage}{10cm}
\centering\vsp
(1)\quad $\cfrac{t_i\ge\rpo u}{f\vt>\rpo u}$\\[2mm]

(2)\quad $\cfrac{f>_\cF g\quad f\vt>\rpo \vu}{f\vt>\rpo g\vu}$\\[2mm]

(3)\quad $\cfrac{f\simeq_\cF g\quad \vt~(>\rpo)\stat{f}~\vu\quad
f\vt>\rpo \vu}{f\vt>\rpo g\vu}$\vsp
\end{minipage}}
\end{figure}


\comment{
\begin{theorem}
${>\rpo}={>\rco}$.
\end{theorem}

\begin{proof}
We first prove that ${>\rco}\sle{>\rpo}$.

\begin{lst}{--}
\item [(arg)] By (1).
\item [(prec)] By induction hypothesis, $f\vt>\rpo \vu$. Thus, by (2),
$f\vt>\rpo g\vu$.
\item [(call)] By induction hypothesis, $f\vt>\rpo \vu$ and
$\vt~(>\rpo)\stat{f}~\vu$. Thus, by (3), $f\vt>\rpo g\vu$.
\item [(red)] By induction hypothesis, $f\vt>\rpo u>\rpo v$.
Thus, by transitivity of $>\rpo$, $f\vt>\rpo v$.
\end{lst}

We now prove that ${>\rpo}\sle{>\rco}$.

\begin{enumi}{}
\item By (arg), $f\vt>\rco t_i$. If $u=t_i$ then we are done.
Otherwise, $t_i>\rpo u$ and, by induction hypothesis,
$t_i>\rco u$. Thus, by (red), $f\vt>\rco u$.
\item By induction hypothesis, $f\vt>\rco \vu$. Thus, by (prec),
$f\vt>\rco g\vu$.
\item By induction hypothesis, $\vt~(>\rco)\stat{f}~\vu$ and $f\vt>\rco
\vu$. Thus, by (call), $f\vt>\rco g\vu$.\qed
\end{enumi}
\end{proof}
}

\section{Preliminaries to the higher-order case}

Before presenting the computability closure for the higher-order case,
we first present the ingredients of the termination proof. As
explained in the introduction, it is based on an adaptation of Tait's
computability technique. First, we interpret each type by a set of
computable terms and prove common properties about computable
terms. Then, following \cite{blanqui04rta}, we define some ordering on
computable terms that will be used in the place of the subterm
ordering for comparing arguments in recursive calls.


We consider simply-typed $\l$-terms with curried constants. Let $\cB$
be a set of {\em base types}. The set $\bT$ of {\em simple types} is
inductively defined as usual. The set $\pos(T)$ of {\em positions in a
type $T$} is defined as usual as words on $\{1,2\}$. The sets
$\pos^+(T)$ and $\pos^-(T)$ of {\em positive and negative positions}
respectively are inductively defined as follows:

\begin{lst}{--}
\item $\pos^\d(\fB)=\{\vep\}$.
\item $\pos^\d(T\A U)= 1\cdot\pos^{-\d}(T)\cup 2\cdot\pos^\d(U)$.
\end{lst}

Let $\pos(\fB,T)$ be the positions of the occurrences of $\fB$ in
$T$. A base type $\fB$ {\em occurs only positively} (resp. {\em
negatively}) in a type $T$ if $\pos(\fB,T)\sle\pos^+(T)$
(resp. $\pos(\fB,T)\sle\pos^-(T)$).


Let $\cX$ be a set of {\em variables} and $\cF$ be a disjoint set of
{\em symbols}. We assume that every $a\in\cX\cup\cF$ is equipped with
a type $T_a\in\bT$. The sets $\cT^T$ of {\em terms of type $T$} are
inductively defined as follows:

\begin{lst}{--}
\item If $a\in\cX\cup\cF$, then $a\in\cT^{T_a}$.
\item If $x\in\cX$ and $t\in\cT^U$, then $\lx t\in\cT^{T_x\A U}$.
\item If $v\in\cT^{T\A U}$ and $t\in\cT^T$, then $vt\in\cT^U$.
\end{lst}

As usual, we assume that, for all type $T$, the set of variables of
type $T$ is infinite, and consider terms up to type-preserving
renaming of bound variables. In the following, $t:T$ or $t^T$ means
that $t\in\cT^T$. Let $\FV(t)$ be the set of variables {\em free} in
$t$.


\begin{definition}[Accessible arguments]
For every $f^{\vT\A\fB}\in\cF$, let $\Acc(f)=
\{i\le|\vT|~|~ \pos(\fB,T_i)\sle\pos^+(T_i)\}$.
\end{definition}


\begin{definition}[Rewrite rules]
A {\em rewrite rule} is a pair of terms $(t^T,u^U)$ such that $t$ is
of the form $f\vt$, $\FV(u)\sle\FV(t)$ and $T=U$.
\end{definition}

In the following, we assume given a set $R$ of rewrite rules. Let
${\a}={{\ab}\cup{\a_R}}$, $\SN=\SN(\a)$ and $\SN^T=\SN\cap\cT^T$. Let
$\cC$ be the set of symbols $c$ such that, for every rule $(f\vt,u)\in
R$, $f\neq c$. The symbols of $\cC$ are said {\em constant}, while the
symbols of $\cD=\cF\moins\cC$ are said {\em defined}.


\subsection{Interpretation of types}

\begin{definition}[Interpretation of types]
A term is {\em neutral\,} if it is of the form $x\vu$ or of the form
$(\lx t)\vu$. Let $\cQ_R^T$ be the set of all sets of terms $P$ such
that:

\begin{enumi}{}
\item $P\sle\SN^T$.
\item $P$ is stable by $\a$.
\item If $t:T$ is neutral and $\red{t}\sle P$, then $t\in P$.
\end{enumi}

\noindent
Let $\cI_R$ be the set of functions $I$ from $\cB$ to
$\bigcup_{\fB\in\cB}\cQ_R^\fB$ such that, for all $\fB\in\cB$,
$I(\fB)\in\cQ_R^\fB$. Given an {\em interpretation of base types}
$I\in\cI_R$, we define an interpretation $\I{T}_R^I\in\cQ_R^T$ for any
type $T$ as follows:

\begin{lst}{--}
\item $\I\fB_R^I=I(\fB)$,
\item $\I{T\A U}_R^I=\{v\in\SN^{T\A U}~|~\all t\in\I{T}_R^I,\,vt\in\I{U}_R^I\}$.
\end{lst}

\noindent
We also let $F_R^I(\fB)=\{t\in\SN^B~|~\all
f^{\vT\A\fB}\vt,\,t\a^* f\vt\A\all i\in\Acc(f),\,t_i\in\I{T_i}_R^I\}$.
\end{definition}

Ordered point-wise by inclusion, $\cI_R$ is a complete lattice.


\comment{
\begin{lemma}
If $I\in\cI_R$ then, for all type $T$, $\I{T}_R^I\in\cQ_R^T$.
\end{lemma}

\begin{proof}
By induction on $T$. This is immediate if $T\in\cB$. Assume now that
$\I{T}_R^I\in\cQ_R^T$ and $\I{U}_R^I\in\cQ_R^U$. We
prove that $\I{T\A U}_R^I\in\cQ_R^{T\A U}$.

\begin{enumi}{}
\item $\I{T\A U}_R^I\sle\SN^{T\A U}$ by definition.
\item Let $v\in\I{T\A U}_R^I$, $v'\in\red{v}$ and $t\in\I{T}_R^I$.
  We must prove that $v't\in\I{U}_R^I$. This follows from the
  facts that $\I{U}_R^I\in\cQ_R^U$, $vt\in\I{U}_R^I$ and
  $v't\in\red{vt}$.
\item Let $v^{T\A U}$ be a neutral term such that
  $\red{v}\sle\I{T\A U}_R^I$ and $t\in\I{T}_R^I$. We must
  prove that $vt\in\I{U}_R^I$. Since $v$ is neutral, $vt$ is
  neutral too. Since $\I{U}_R^I\in\cQ_R^U$, it suffices to
  prove that $\red{vt}\sle\I{U}_R^I$. Since
  $\I{T}_R^I\in\cQ_R^T$, $t\in\SN$ and we can proceed by
  induction on $t$ with $\a$ as well-founded ordering. Let
  $w\in\red{vt}$. Since $v$ is neutral, either $w=v't$ with
  $v'\in\red{v}$, or $w=vt'$ with $t'\in\red{t}$. In the former case,
  $w\in\I{U}_R^I$ since $v'\in\I{T\A U}_R^I$. In the latter
  case, we conclude by induction hypothesis on $t'$.\qed
\end{enumi}
\end{proof}
}


\begin{lemma}
$F_R$ is a monotone function on $\cI_R$.
\end{lemma}

\begin{proof}
We first prove that $P=F_R^I(\fB)\in\cQ_R^\fB$.

\begin{enumi}{}
\item $P\sle\SN^\fB$ by definition.
\item Let $t\in P$, $t'\in\red{v}$, $f:\vT\A\fB$ and $\vt$
  such that $t'\a^* f\vt$. We must prove that
  $\vt\in\I{\vT}_R$. It follows from the facts that $t\in P$ and
  $t\a^* f\vt$.
\item Let $t^\fB$ neutral such that $\red{t}\sle P$.
  Let $f^{\vT\A\fB}$, $\vt$ such that $t\a^* f\vt$ and
  $i\in\Acc(f)$. We must prove that $t_i\in\I{T_i}_R$. Since $t$ is
  neutral, $t\neq f\vt$. Thus, there is $t'\in\red{t}$ such that
  $t'\a^* f\vt$. Since $t'\in P$, $t_i\in\I{T_i}_R$.
\end{enumi}

For the monotony, let ${\le^+}={\le}$ and ${\le^-}={\ge}$. Let $I\le
J$ iff, for all $\fB$, $I(\fB)\sle J(\fB)$. We first prove that
$\I{T}_R^I\sle^\d \I{T}_R^J$ whenever $I\le J$ and
$\pos(\fB,T)\sle\pos^\d(T)$, by induction on $T$.

\begin{lst}{--}
\item Assume that $T=C\in\cB$. Then, $\d=+$, $\I{T}_R^I=I(C)$
  and $\I{T}_R^I=J(C)$. Since $I(C)\sle J(C)$, $\I{T}_R^I\sle
  \I{T}_R^I$.
\item Assume that $T=U\A V$. Then, $\pos(\fB,U)\sle\pos^{-\d}(U)$ and
  $\pos(\fB,V)\sle\pos^\d(V)$. Thus, by induction hypothesis,
  $\I{U}_R^I\sle^{-\d} \I{U}_R^J$ and $\I{V}_R^I\sle^\d
  \I{V}_R^J$. Assume that $\d=+$. Let $t\in\I{T}_R^I$ and
  $u\in\I{U}_R^J$. We must prove that $tu\in\I{V}_R^J$. Since
  $\I{U}_R^I\sge \I{U}_R^J$, $tu\in\I{V}_R^I$. Since $\I{V}_R^I\sle
  \I{V}_R^J$, $tu\in\I{V}_R^J$. It works similarly for $\d=-$.
\end{lst}

Assume now that $I\le J$. We must prove that, for all $\fB$,
$F_R^I(\fB)\sle F_R^J(\fB)$. Let $\fB\in\cB$ and $t\in F_R^I(\fB)$. We must
prove that $t\in F_R^J(\fB)$. First, we have $t\in\SN^\fB$ since $t\in
F_R^I(\fB)$. Assume now that $t\a^* f^{\vT\A\fB}\vt$ and let
$i\in\Acc(f)$. We must prove that $t_i\in\I{T_i}_R^J$. Since $t\in
F_R^I(\fB)$, $t_i\in\I{T_i}_R^I$. Since $i\in\Acc(f)$,
$\pos(\fB,T_i)\sle\pos^+(T_i)$ and $\I{T_i}_R^I\sle \I{T_i}_R^J$.\qed
\end{proof}


\begin{definition}[Computability]
Let $I_R$ be the least fixpoint of $F_R$. A term $t:T$ is {\em
$R$-computable} if $t\in \I{T}_R= \I{T}_R^{I_R}$.
\end{definition}


\subsection{Computability properties}

\begin{lemma}
\label{lem-comp-lam}
If $t$, $u$ and $t\xu$ are computable, then $(\lx t)u$ is computable.
\end{lemma}

\begin{proof}
Since $(\lx t)u$ is neutral, it suffices to prove that every reduct is
computable. Since $t$ and $u$ are $\SN$, we can proceed by induction
on $(t,u)$ with $\a\lex$ as well-founded ordering. Assume that $(\lx
t)u\a v$. If $v=t\xu$, then $t'$ is computable by
assumption. Otherwise, $v=(\lx t')u$ with $t\a t'$, or $v=(\lx t)u'$
with $u\a u'$. In both cases, we can conclude by induction
hypothesis.\qed
\end{proof}


\begin{lemma}
\label{lem-comp-symb}
A term $f\vt:\fB$ is computable whenever every reduct of $f\vt$ is
computable and, for all $i\in\Acc(f)$, $t_i$ is computable.
\end{lemma}

\begin{proof}
Assume that $f\vt\a^* g\vu$ with $g:\vU\A\fB$. Let $i\in\Acc(g)$. If
$f\vt\neq g\vu$, then there is $v\in\red{f\vt}$ such that $v\a^*
g\vu$. Since $v$ is computable, $u_i$ is computable. Otherwise,
$u_i=t_i$ is computable by assumption.\qed
\end{proof}


\begin{lemma}
\label{lem-comp-const}
Every constant symbol is computable.
\end{lemma}

\begin{proof}
Let $c^{\vT\A\fB}\in\cC$ and $\vt\in\I{\vT}_R$. By Lemma
\ref{lem-comp-symb}, $c\vt$ is computable if every reduct of $c\vt$ is
computable. Since $\vt\in\SN$, we can proceed by induction on $\vt$
with $\a\lex$ as well-founded ordering. Assume that $c\vt\a u$. Since
$c\in\cC$, $u=c\vt'$ with $\vt\a\lex\vt'$. Thus, by induction
hypothesis, $c\vt'$ is computable.\qed
\end{proof}


\begin{lemma}
\label{lem-comp}
If every defined symbol is computable, then every term is computable.
\end{lemma}

\begin{proof}
First note that the identity substitution is computable since
variables are computable (they are neutral and irreducible). We then
prove that, for every term $t$ and computable substitution $\t$, $t\t$
is computable, by induction on $t$.

\begin{lst}{--}
\item Assume that $t=f\in\cD$. Then, by assumption, $t\t=f$ is computable.
\item Assume that $t=c\in\cC$. Then, by Lemma \ref{lem-comp-const},
$t\t=c$ is computable.
\item Assume that $t=x\in\cX$. Then, $t\t=x\t$ is computable since
  $\t$ is computable.
\item Assume that $t=\lx u$. Then, $t\t= \lx u\t$.
  Let $v\in\I{V}_R$. We must prove that $t\t v\in\I{U}_R$. By
  induction hypothesis, $u\t\xv$ is computable. Since $u\t$ and $v$
  are computable too, by Lemma \ref{lem-comp-lam}, $t\t$ is
  computable.
\item Assume that $t=u^{V\A T}v$. Then, $t\t=u\t v\t$. By induction
  hypothesis, $u\t\in\I{V\A T}_R$ and $v\t\in\I{V}_R$. Thus, $t\t\in
  \I{T}_R$.\qed
\end{lst}
\end{proof}


\subsection{Size ordering}

The least fixpoint of $F_R$, $I_R$, is reachable by transfinite
iteration from the smallest element of $\cI_R$. This provides us with
the following ordering.

\begin{definition}[Size ordering]
For all $\fB\in\cB$ and $t\in\I\fB_R$, let the {\em size} of $t$ be the
smallest ordinal $o_R^\fB(t)=\ka$ such that $t\in F_R^\ka(\vide)(\fB)$,
where $F_R^\ka$ is the transfinite $\ka$-iteration of $F_R$. Let
$\succ_R=\bigcup_{T\in\cB^\A}\succ_R^T$, where $(\succ_R^T)_{T\in\cB^\A}$
is the family of orderings inductively defined as follows:

\begin{lst}{--}
\item For all $\fB\in\cB$, let $t\succ_R^\fB u$ iff $t,u\in\I\fB_R$
  and $o_R^\fB(t)>o_R^\fB(u)$.
\item For all $T,U\in\cB^\A$, let $t\succ_R^{T\A U} u$ iff $t,u\in\I{T\A U}_R$
  and, for all $v\in\I{T}_R$, $tv\succ_R^U uv$. 
\end{lst}
\end{definition}

In the first-order case, recursive call arguments where compared with
the subterm ordering. But the subterm ordering is not adapted to
higher-order rewriting. Consider for instance the following
simplification rule on process algebra \cite{vandepol93hoa}:

\begin{center}
$(\S P);x\a \S(\ly Py;x)$
\end{center}

\noindent
where $\S^{(\fD\A\fP)\A\fP}$ is a data-dependent choice operator and
$;^{\fP\A\fP\A\fP}$ the sequence operator. The term $Py$ is not a
subterm of $\S P$. The interpretation of $P$ gives us the solution:
$\I\fP_R=\{t\in\SN^P~|~\all f^{\vT\A\fP}\vt,\,t\a^* f\vt\A\all
i\in\Acc(f),\,t_i\in\I{T_i}_R\}$. Since $\fP$ occurs only positively
in $\fD\A\fP$, $\Acc(\S)=\{1\}$. Hence, if $\S P\in\I\fP_R$ then, for
all $d\in\I\fD_R$, $pd\in\I\fP_R$ and $o_R^\fP(Pd)<o_R^\fP(\S P)$.


We immediately check that the size ordering is well-founded.

\begin{lemma}
$\succ_R^T$ is transitive and well-founded.
\end{lemma}

\begin{proof}
By induction on $T$. For $T\in\cB$, this is immediate. Assume now that
$(t_i)_{i\in\bN}$ is an increasing sequence for $\succ_R^{T\A
U}$. Since variables are computable, let $x\in\I{T}_R$. By definition
of $\succ_R^{T\A U}$, $(t_ix)_{i\in\bN}$ is an increasing sequence for
$\succ_R^U$.\qed
\end{proof}


In case of a first-order type $\fB$, when $\a$ is confluent, the size
of $t^\fB$ is the number of (constructor) symbols at the top of its
normal form. So, it is equivalent to using embedding on normal
forms. But, since the ordering is compatible with reduction, in the
sense that $t\succeq_R u$ whenever $t\a u$, it is finer than the
embedding. For instance, by taking the rules:

\begin{rewc}
x-0 & x\\
0-x & 0\\
(sx)-(sy) & x-y\\
\end{rewc}

\noindent
one can prove that $t-u\tle_R t$. This allows to prove the termination
of functions for which simplification orderings fail like:

\begin{rewc}
0/y & 0\\
(sx)/y & s((x-y)/y)\\
\end{rewc}


However, in practice, the size ordering cannot be used as is. We need
a decidable syntactic approximation. In \cite{blanqui04rta}, we assume
given an ordered term algebra $(\cA,>_\cA)$ for representing
operations on ordinals and, for each base type $\fB$ and expression
$a\in\cA$, we introduce the subtype $\fB^a$ of terms of type $\fB$ whose
size is less than or equal to $a$. Then, in the (call) rule, the size
annotations of $\vt$ and $\vu$ are compared with $>_\cA$. In
\cite{blanqui05csl}, we prove that type checking is decidable, whenever
the constraints generated by these comparisons are satisfiable, hence
providing a powerful termination criterion. We do not use size
annotations here, but it would definitely be a natural and powerful
extension. Instead, we are going to define an approximation like in
\cite{blanqui05mscs}.


\section{Higher-order case}

We now introduce the size-ordering approximation and the computability
closure for the higher-order case.


\begin{figure}[ht]
\centering\caption{Higher-order computability closure\label{fig-hocc}}\vsp
\fbox{\begin{minipage}{11cm}
\centering\vsp
(arg)\quad $t_i\in\CC_R^f(\vt)$\\[2mm]

(decomp)\quad $\cfrac{g\vu\in\CC_R^f(\vt)\quad i\in\Acc(g)}
{u_i\in\CC_R^f(\vt)}$\\[2mm]

(prec)\quad $\cfrac{f>_\cF g}
{g\in\CC_R^f(\vt)}$\\[2mm]

(call)\quad $\cfrac{f\simeq_\cF g^{\vU\A U}\quad
\vu^\vU\in\CC_R^f(\vt)\quad \vt~({\a_{\b R}^+}\cup{\tgt\!_R^{f\vt}\,})\stat{f}~\vu}
{g\vu\in\CC_R^f(\vt)}$\\[2mm]

(red)\quad $\cfrac{u\in\CC_R^f(\vt)\quad u\a_{\b R}^+ v}
{v\in\CC_R^f(\vt)}$\\[2mm]

(app)\quad $\cfrac{u^{V\A T}\in\CC_R^f(\vt)\quad
v^V\in\CC_R^f(\vt)}{uv\in\CC_R^f(\vt)}$\\[2mm]

(var)\quad $\cfrac{x\notin\FV(\vt)}{x\in\CC_R^f(\vt)}$\\[2mm]

(lam)\quad $\cfrac{u\in\CC_R^f(\vt)\quad x\notin\FV(\vt)}
{\lx u\in\CC_R^f(\vt)}$\vsp
\end{minipage}}
\end{figure}


\begin{figure}[ht]
\centering\caption{Ordering for comparing function arguments\label{fig-comp-ord}}\vsp
\fbox{\begin{minipage}{11cm}
\centering\vsp
($\tgt$base)\quad $\cfrac{i\in\Acc(g)\quad \vb\in\CC_R^f(\vt)}
{g^{\vA\A B}\va^\vA\tgt\!_R^{f\vt}\, a_i^{\vB\A B}\vb^\vB}$\\[2mm]

($\tgt$lam)\quad $\cfrac{a\tgt\!_R^{f\vt}\, bx\quad x\notin\FV(b)\cup\FV(\vt)}
{\lx a\tgt\!_R^{f\vt}\, b}$\\[2mm]

($\tgt$red)\quad $\cfrac{a\tgt\!_R^{f\vt}\, b\quad b\a_{\b R}^+ c}
{a\tgt\!_R^{f\vt}\, c}$\\[2mm]

($\tgt$trans)\quad $\cfrac{a\tgt\!_R^{f\vt}\, b\quad b\tgt\!_R^{f\vt}\, c}
{a\tgt\!_R^{f\vt}\, c}$\vsp
\end{minipage}}
\end{figure}


\begin{definition}[Computability closure]
The {\em computability closure} of a term $f\vt$, written
$\CC_R^f(\vt)$, and the associated size-ordering approximation,
written $\tgt\!_R^{f\vt}$, are mutually inductively defined in Figures
\ref{fig-comp-ord} and \ref{fig-hocc} respectively. Let $\CR(R)$ be
the set of pairs $(f\vt,u)$ such that $u\in\CC_R^f(\vt)$,
$\FV(u)\sle\FV(f\vt)$ and $f\vt$ and $u$ have the same type.
\end{definition}

Compared to the first-order case, we added the rules (var) and (lam)
to build abstractions and, in (call), we replaced $\a_R^+$ by $\a_{\b
R}^+$, and $\tgt$ by $\tgt\!_R^{f\vt}$. This ordering is a better
approximation of the size ordering than the one given in
\cite{blanqui05mscs} where, in ($\tgt$base),
$\vb\in\cX\moins\FV(\vt)$. In this case, the size-ordering
approximation can be defined independently of the computability
closure. Note however that, in both cases, the size-ordering
approximation contains the subterms of same type. In the process
algebra example, by ($\tgt$base), we have $\S P\tgt\!_R^l\,Py$ where
$l=(\S P);x$.

We now prove the correctness of the computability closure.


\begin{lemma}
\label{lem-hocr-wf}
If ${R}\sle{\CR(R)}$, then ${\ab}\cup{\a_{\CR(R)}}$ is well-founded.
\end{lemma}

\begin{proof}
Let $S=\CR(R)$. It suffices to prove that every term is
$S$-computable. Let ${\a}={{\ab}\cup{\a_S}}$ and $\SN=\SN(\a)$. After
Lemma \ref{lem-comp}, it suffices to prove that, for all $f^{\vV\A B}$
and $\vv\in\I{\vV}_S$, $f\vv\in\I{B}_S$. We prove it by induction on
$((f,\vv),\vv)$ with $((\succ_S)\stat{},\a\lex)$ as well-founded ordering
($\vv$ are computable) (H1). By Lemma \ref{lem-comp-symb}, it suffices
to prove that $\red{f\vv}\sle\I{B}_S$. Let $v'\in\red{f\vv}$. Either
$v'=f\vv'$ with $\vv\a\stat{f}\vv'$, or $v=f\vt\s$, $v'=u\s$ and
$u\in\CC_R^f(\vt)$. In the former case, $\vv'\in\I{\vV}_S$ since
$\I{\vV}_S$ is stable by $\a$, and $\vv(\tge_S)\stat{f}\vv'$. Thus, we
can conclude by (H1). For the latter case, we prove that, if
$u\in\CC_R^f(\vt)$ then, for all $S$-computable substitution $\t$ such
that $\dom(\t)\sle\FV(u)\moins\FV(\vt)$, $u\s\t$ is $S$-computable, by
induction on $\CC_R^f(\vt)$ (H2).

\begin{lst}{--}
\item [(arg)] $t_i\s=v_i$ is computable by assumption.

\item [(decomp)] By (H2), $g\vu\s\t$ is computable.
  Thus, by definition of $I_S$, $u_i\s\t$ is computable.

\item [(prec)] By (H1), $g$ is computable.

\item [(call)] By (H2), $\vu\s\t$ are computable.
  Since $\dom(\t)\cap\FV(\vt)=\vide$, $t_i\s\t=t_i\s=v_i$. Assume that
  $t_i\a_{\b R}^+ u_j$. Then, $v_i\a_{\b R}^+ u_j\s\t$. Since
  ${R}\sle{S}$ and ${\a_{\b S}^+}\sle{\tge_S}$, ${v_i}\tge_S
  {u_j\s\t}$. Assume now that $t_i\tgt\!_R^f u_j$. We prove that, if
  $a\tgt\!_R^{f\vt}\, b$ then, for all $S$-computable substitution $\t$
  such that $\dom(\t)\sle\FV(b)\moins(\FV(a)\cup\FV(\vt))$ and $a\s\t$
  is $S$-computable, $b\s\t$ is $S$-computable and $a\s\t\succ_S
  b\s\t$.

\begin{lst}{--}
\item [($\tgt$base)] Let $a=g^{\vA\A B}\va^\vA$ and $b=a_i^{\vB\A B}\vb^\vB$.
  Let $I_S^\ka=F_S^\ka(\vide)$. Note that the size of a term is
  necessarily a successor ordinal. Thus, $o_S(a\s\t)=\ka+1$ and, by
  definition of $\I{B}_S$, $a_i\s\t\in\I{\vB\A B}_S^{I_S^\ka}$. Since
  $\vb\in\CC_R^f(\vt)$ and $\dom(\t)\sle\FV(\vb)\moins\FV(\vt)$, by
  (H2), $\vb\s\t$ are computable. Therefore, $a_i\s\t\vb\s\t\in
  I_S^\ka(B)$ and $o_S(b\s\t)\le\ka<o_S(a\s\t)$.

\item [($\tgt$lam)] Let $w\in\I{T_x}_S$. We must prove that $b\s\t w$ is
  computable. Since $x\notin\FV(b)\cup\FV(\vt)$,
  $x\notin\dom(\s\t)$. W.l.o.g., we can assume that
  $x\notin\codom(\s\t)$. Thus, $(\lx a)\s\t=\lx a\s\t$. Let
  $\t'=\t\cup\xw$. Since $\lx a\s\t$ is computable, $a\s\t'$ is
  computable. Since $\dom(\t')\sle\FV(bx)\moins(\FV(a)\cup\FV(\vt))$,
  by induction hypothesis, $(bx)\s\t'=b\s\t' w$ is computable and
  $a\s\t'\succ_S b\s\t' w$. Since $x\notin\dom(\s)$,
  $b\s\t'=b\s\t$. Thus, $b\s\t$ is computable and $(\lx a)\s\t\succ_S
  b\s\t$.

\item [($\tgt$red)] By induction hypothesis and since
  ${\a_{\b S}^+}\sle{\tge_S}$.

\item [($\tgt$trans)] By induction hypothesis and transitivity of $\succ_S$.
\end{lst}

  Hence, $v_i=t_i\s\t\succ_S u_j\s\t$ since
  $\dom(\t)\sle\FV(u_j)\moins(\FV(t_i)\cup\FV(\vt))$ and $v_i$ is
  computable. Therefore, either $\vv(\succ_S)\stat{f}\vu\s\t$ or
  $\vv\a^+\stat{f}\vu\s\t$ and, by (H1), $f\vu\s\t$ is computable.

\item [(red)] By (H2), $u\s\t\in\I{U}_S$.
  Since $\a_{\b R}^+$ is stable by substitution, $u\s\t\a_{\b R}^+
  v\s\t$. Since ${R}\sle{S}$, $u\s\t\a^+ v\s\t$. Since $\I{U}_S$ is
  stable by $\a$, $v\s\t$ is computable.

\item [(app)] By (H1), $u\s\t$ and $v\s\t$ are computable.
  Thus, by definition of $\I{V\A T}_S$, $u\s\t v\s\t$ is computable.

\item [(lam)] W.l.o.g, we can assume that
  $x\notin\dom(\t)\cup\codom(\s\t)$. Thus, $(\lx u)\s\t= \lx
  u\s\t$. Let $v:T_x$ computable and $\t'=\t\cup\xv$. If
  $x\notin\FV(u)$, then $u\s\t'=u\s\t$ is computable. Otherwise, since
  $\dom(\t')= \dom(\t)\cup\{x\}$, $\dom(\t)\sle \FV(\lx
  u)\moins\FV(\vt)$ and $x\notin\FV(\vt)$, we have $\dom(\t')\sle
  \FV(u)\moins\FV(\vt)$. Thus, by (H2), $u\t'$ is computable. Hence,
  by Lemma \ref{lem-comp-lam}, $\lx u\t$ is computable.

\item [(var)] Since $x\notin\FV(\vt)$, $x\s\t=x\t$ is computable
  by assumption on $\t$.\qed
\end{lst}
\end{proof}


Like in the first-order case, one can easily check that the functions
$\tgt\!^{f\vt}$, $\CC^f(\vt)$ and $\CR$ are monotone and
$\w$-sup-con\-tinuous for inclusion.

\begin{definition}[Higher-order recursive computability ordering]
Let the {\em weak higher-order recursive computability ordering}
$>\whorco$ be the least fixpoint of $\CR$, and the {\em higher-order
recursive computability ordering} $>\horco$ be the closure by context
of $>\whorco$.
\end{definition}


In the following, let ${\tgt\whorco}={\tgt_{>\whorco}}$ and
$\CC=\CC_{>\whorco}$. The well-foundedness of $\ab\cup>\horco$
immediately follows from Lemma \ref{lem-hocr-wf} and the facts that
${>\whorco}\sle{\CR(>\whorco)}$ and ${\a_{>\whorco}}={>\horco}$.

\begin{theorem}
$\ab\cup>\horco$ is a well-founded rewrite relation.
\end{theorem}


Before comparing $>\horco$ with the monomorphic version of $>\horpo$
\cite{jouannaud99lics} whose definition is recalled in Figure
\ref{fig-horpo}, let us give some examples.

\begin{example}[Differentiation]
Taken from \cite{blanqui02tcs} (Example 10 in
\cite{jouannaud01draft}). Consider the symbols $0^\fR$, $1^\fR$,
$+^{\fR\A\fR\A\fR}$, $\times^{\fR\A\fR\A\fR}$, and
$D^{(\fR\A\fR)\A\fR\A\fR}$. The rule:

\begin{center}
$D\lx Fx\times Gx\a \lx DFx\times Gx+ Fx\times DGx$
\end{center}

\noindent
is both in $>\horco$ and $>\horpo$. Take $D>_\cF\times,+$. By (prec),
$t=D\lx Fx\times Gx>+,\times$. By (var), $t>x$. By (arg), $t>\lx
Fx\times Gx$. By (app), $t>(\lx Fx\times Gx)x$. By (red), $t>Fx\times
Gx$. Since $\Acc(\times)=\{1,2\}$, by (decomp), $t>Fx,Gx$. By
($\tgt$base), $Fx\times Gx\tgt Fx,Gx$. By ($\tgt$lam), $\lx
Fx\times Gx\tgt F,G$. By (call), $t>DF,DG$. By several applications
of (app), $t>DFx\times Gx+ Fx\times DGx$. Finally, by (abs), $t>\lx
DFx\times Gx+ Fx\times DGx$.
\end{example}


We now give two examples included in $>\horco$ but not in $>\horpo$.

\begin{example}[Process Algebra]
Taken from \cite{vandepol93hoa} (Example 5 in
\cite{jouannaud99lics}). The rule:

\begin{center}
$(\S P);x\a \S(\ly Py;x)$
\end{center}

\noindent
is in $>\horco$ but not in $>\horpo$. Take ${\S}<_\cF{;}$ and
$\mr{stat}_;=\mr{lex}$. By (arg), $t=(\S P);x>\S P,x$. Since
$\Acc(\S)=\{1\}$, by (decomp), $t>P$. By (var), $t>y$. By (app),
$t>Py$. By ($\tgt$base), $\S P\tgt Py$. By (call), $t>Py;x$. By (lam),
$t>\ly Py;x$. Thus, by (prec), $t>\S\ly Py;x$.
\end{example}


\begin{example}[Lists of functions]
This is Example 6 in \cite{jouannaud99lics}. Consider the symbols
$\mi{fcons}^{(\fB\A\fB)\A\fL\A\fL}$ and
$\mi{lapply}^{\fB\A\fL\A\fB}$. The rule:

\begin{center}
$\mi{lapply}\,x\,(\mi{fcons}\,F\,l)\a F\,(\mi{lapply}\,x\,l)$
\end{center}

\noindent
is in $>\horco$ but not in $>\horpo$. Take
$\mr{stat}_\mi{lapply}=\mr{lex}$ (from right to left). By (arg),
$t=\mi{lapply}\,x\,(\mi{fcons}\,F\,l)>x,\mi{fcons}\,F\,l$. Since
$\Acc(\mi{fcons})=\{1,2\}$, by (decomp), $t>F,l$. By ($\tgt$base),
$\mi{fcons}\,F\,l\tgt l$. By (call), $t>\mi{lapply}\,x\,l$. Thus, by
(app), $t>F\,(\mi{lapply}\,x\,l)$.
\end{example}


\section{Comparison with HORPO}

Before proving that ${>\horpo}\sle{>\horco^+}$, we study some
properties of $>\horco$.

\begin{lemma}
\label{lem-hoco-prop}
\begin{enumi}{}
\comment{
\item\label{itema} If $t\tgt\whorco u$, $\dom(\t)\sle\FV(t)$
and $\codom(\t)\cap(\FV(u)\moins\FV(t))=\vide$, then $t\t\tgt\whorco u\t$.
\item\label{itemb} If $t\tgt\whorco u$ and $(\FV(u)\moins\FV(t))\cap\FV(w)=\vide$,
then $tw\tgt\whorco u$.
}
\item\label{itemc} $>\whorco$ is stable by substitution.
\item\label{itemd} ${>\whorco\a^+}\sle{>\whorco}$.
\item\label{iteme} If $t>\whorco u$, then $t\vw>\whorco u\vw$.
\item\label{itemf} If $t\a^+ u$, then $f\va t\vb>\whorco f\va u\vb$.
\item\label{itemg} $>\whorco$ is transitive.
\item\label{itemh} ${>\horco^+>\whorco}\sle{>\whorco}$.
\end{enumi}
\end{lemma}

From (\ref{itemd}) and (\ref{itemh}), it follows that any sequence of
$>\horco$-steps with at least one $>\whorco$-step, in fact corresponds
to a $>\whorco$-step. So, $>\horco$ is not far from being transitive.


\begin{figure}[ht]
\centering\caption{HORPO \cite{jouannaud99lics}\label{fig-horpo}}\vsp
\fbox{\begin{minipage}{11cm}
\centering\vsp
$P(f,\vt,u)= f\vt>\horpo u\ou (\ex j)~ t_j\ge\horpo u$
\vsp

(1)\quad $\cfrac{t_i\ge\horpo u}{f^{\vT\A T}\vt^\vT>\horpo u^T}$\\[2mm]

(2)\quad $\cfrac{f>_\cF g\quad P(f,\vt,\vu)}
{f^{\vT\A T}\vt^\vT>\horpo g^{\vU\A T}\vu^\vU}$\\[2mm]

(3)\quad $\cfrac{f\simeq_\cF g\quad
\mr{stat}_f=\mr{mul}\quad \vt~(>\horpo)\stat{f}~\vu}
{f^{\vT\A T}\vt^\vT>\horpo g^{\vU\A T}\vu^\vU}$\\[2mm]

(4)\quad $\cfrac{f\simeq_\cF g\quad \mr{stat}_f=\mr{lex}\quad
\vt~(>\horpo)\stat{f}~\vu\quad P(f,\vt,\vu)}
{f^{\vT\A T}\vt^\vT>\horpo g^{\vU\A T}\vu^\vU}$\\[2mm]

(5)\quad $\cfrac{P(f,\vt,\vu)}{f^{\vT\A T}\vt>\horpo \vu^T}$\\[2mm]

(6)\quad $\cfrac{\{t_1,t_2\}~(>\horpo)\mul~ \{u_1,u_2\}}
{t_1^{U\A T}t_2^U> u_1^{V\A T}u_2^V}$\\[2mm]

(7)\quad $\cfrac{t>\horpo u}{\lx t>\horpo \lx u}$\vsp
\end{minipage}}
\end{figure}


We now compare $>\horco$ with the monomorphic version of $>\horpo$
defined in Figure \ref{fig-horpo}. For the case (6), let us list all
the cases that may be possible {\em a priori\,}:

\begin{enumalphai}
\item $t_1\ge\horpo u_1$ and $t_1\ge\horpo u_2$.
This case is not possible since then we would have ${U\A T}={V\A T}=V$.

\item $t_2\ge\horpo u_1$ and $t_2\ge\horpo u_2$.
This case is not possible since then we would have $U={V\A T}=V$.

\item $t_1\ge\horpo u_1$ and $t_2\ge\horpo u_2$.
This case is possible.

\item $t_2\ge\horpo u_1$ and $t_1\ge\horpo u_2$.
This case is not possible since then we would have $U={V\A T}$ and
${U\A T}=V$, and thus $U={(U\A T)\A T}$.
\end{enumalphai}

Hence, only (c) is in fact possible. We now prove that
${>\horpo}\sle{>\horco^+}$.


\begin{theorem}
${>\horpo}\sle{>\horco^+}$.
\end{theorem}

\begin{proof}
We first prove that $f\vt>v$ whenever $f\vt>\horco^+ v$ or $t_j>\horco^* v$
(*). Assume that $t_j>\horco^* v$. By (arg), $f\vt>t_j$. Thus, by (red),
$f\vt> v$. Assume now that $f\vt>\horco u>\horco^* v$. There are 2 cases:

\begin{lst}{--}
\item $f\vt=f\va t_k\vb$, $u=f\va t_k'\vb$ and $t_k>\horco t_k'$.
By Lemma \ref{lem-hoco-prop} (\ref{itemf}), $f\vt>\whorco u$. By Lemma
\ref{lem-hoco-prop} (\ref{itemd}), $f\vt>\whorco v$. Thus, $f\vt> v$.

\item $f\vt=f\vl\s\vb$, $u=r\s\vb$ and $f\vl\s>\whorco r\s$.
By Lemma \ref{lem-hoco-prop} (\ref{iteme}), $f\vt>\whorco u$. By Lemma
\ref{lem-hoco-prop} (\ref{itemd}), $f\vt>\whorco v$. Thus, $f\vt> v$.
\end{lst}

We now prove the theorem by induction on $>\horpo$.

\begin{enumi}{}
\item By induction hypothesis, $t_i>\horco^* u$. By (arg), $f\vt> t_i$.
Since $t_i>\horpo u$ and $f\vt>\horpo u$, $(f\vt,t_i)$ is a
rule. Thus, $f\vt>\whorco t_i$ and, by Lemma \ref{lem-hoco-prop}
(\ref{itemd}), $f\vt>\whorco u$.

\item By induction hypothesis, for all $i$, $f\vt>\horco^+ u_i$ or
$t_j>\horco^* u_i$. Hence, by (*), $f\vt>\vu$. By (prec), $f\vt>g$. Thus,
by (app), $f\vt>g\vu$. Since $(f\vt,g\vu)$ is a rule, $f\vt>\whorco g\vu$.

\item By induction hypothesis, $\vt~(>\horco^+)\mul~\vu$.
Hence, by (*), $f\vt>\vu$. Thus, by (call), $f\vt>g\vu$. Since
$(f\vt,g\vu)$ is a rule, $f\vt>\whorco g\vu$.

\item By induction hypothesis, $\vt~(>\horco^+)\stat{f}~\vu$ and,
for all $i$, $f\vt>\horco^+ u_i$ or $t_j>\horco^* u_i$. Hence, by (*),
$f\vt>\vu$. Thus, by (call), $f\vt>g\vu$. Since $(f\vt,g\vu)$ is a
rule, $f\vt>\whorco g\vu$.

\item By induction hypothesis, for all $i$, $f\vt>\horco^+ u_i$ or
$t_j>\horco^* u_i$. Hence, by (*), $f\vt>u_i$ for all $i$. Thus, by (app),
$f\vt>\vu$.  Since $(f\vt,\vu)$ is a rule, $f\vt>\whorco\vu$.

\item As previously remarked, $t_1\ge\horpo u_1$ and $t_2\ge\horpo u_2$.
Thus, by induction hypothesis, $t_1>\horco^* u_1$ and $t_2>\horco^* u_2$.
Hence, by monotony, $t_1t_2>\horco^* u_1t_2>\horco^* u_1u_2$.

\item By induction hypothesis, $t>\horco u$. Thus, by context,
$\lx t>\horco \lx u$.\qed
\end{enumi}
\end{proof}

From the proof, we observe that, if (6) were restricted to
$(t_1>\horpo u_1\et t_2=u_2)\ou (t_1=u_1\et t_2>\horpo u_2)$, then we
would get ${>\horpo}\sle{>\horco}$, since this is the only case requiring
transitivity.

In \cite{jouannaud99lics}, the authors strengthen their definition of
HORPO by adding in $P(f,\vt,\vu)$ the case $u_i\in\cC\cC(f\vt)$, where
$\cC\cC(f\vt)$ is similar to $\CC_\vide^f(\vt)$ with the subterm
ordering $\tgt$ instead of $\tgt\!^f$ in (call). Thus, (*) is still
satisfied and ${>\horpo}\sle{>\horco^+}$ in this case too.

In \cite{jouannaud01draft}, the authors add a few new cases to HORPO
and extend the computability closure a little bit. But, again, this
does not make any essential difference. And, indeed, they recognize
they are not satisfied with their treatment of abstractions. Taking
our interpretation of base types solve these problems.


\section{Conclusion}

We proved that the recursive path ordering is strictly included (equal
in the first-order case) to the recursive computability ordering, an
ordering naturally defined from the notion of computability
closure. In the higher-order case, this does not provide us with a
very practical definition. However, the well-foundedness proof is
reduced to proving the correctness of the computability closure. This
therefore provides us with a way to easily extend HORPO to richer type
systems. For instance, in \cite{blanqui05mscs}, we proved the
correctness of the computability closure for a polymorphic and
dependent type system with both object and type level rewriting. This
would generalize Walukiewicz' extension of HORPO
\cite{walukiewicz03jfp}. In \cite{blanqui03tlca}, we defined an
extension of the computability closure accepting non-simply
terminating systems. Finally, in \cite{blanqui03rta}, we proved that
the computability closure proves the termination of rewriting modulo
AC as well.


\begin{thebibliography}{10}

\bibitem{blanqui05csl}
F.~Blanqui.
\newblock Decidability of type-checking in the {C}alculus of {A}lgebraic
  {C}onstructions with size annotations.
\newblock In {\em Proc. of CSL'05\em, LNCS 3634}.

\bibitem{blanqui01lics}
F.~Blanqui.
\newblock Definitions by rewriting in the {C}alculus of {C}onstructions
  (extended abstract).
\newblock In {\em Proc. of LICS'01}.

\bibitem{blanqui03tlca}
F.~Blanqui.
\newblock Inductive types in the {C}alculus of {A}lgebraic {C}onstructions.
\newblock In {\em Proc. of TLCA'03\em, LNCS 2701}.

\bibitem{blanqui03rta}
F.~Blanqui.
\newblock Rewriting modulo in {D}eduction modulo.
\newblock In {\em Proc. of RTA'03\em, LNCS 2706}.

\bibitem{blanqui00rta}
F.~Blanqui.
\newblock Termination and confluence of higher-order rewrite systems.
\newblock In {\em Proc. of RTA'00\em, LNCS 1833}.

\bibitem{blanqui04rta}
F.~Blanqui.
\newblock A type-based termination criterion for dependently-typed higher-order
  rewrite systems.
\newblock In {\em Proc. of RTA'04\em, LNCS 3091}.

\bibitem{blanqui05mscs}
F.~Blanqui.
\newblock Definitions by rewriting in the {C}alculus of {C}onstructions.
\newblock {\em Mathematical Structures in Computer Science}, 15(1):37--92,
  2005.

\bibitem{blanqui99rta}
F.~Blanqui, J.-P. Jouannaud, and M.~Okada.
\newblock The {C}alculus of {A}lgebraic {C}onstructions.
\newblock In {\em Proc. of RTA'99\em, LNCS 1631}.

\bibitem{blanqui02tcs}
F.~Blanqui, J.-P. Jouannaud, and M.~Okada.
\newblock Inductive-data-type {S}ystems.
\newblock {\em Theoretical Computer Science}, 272:41--68, 2002.

\bibitem{borralleras01lpar}
C.~Borralleras and A.~Rubio.
\newblock A monotonic higher-order semantic path ordering.
\newblock In {\em Proc. of LPAR'01\em, LNCS 2250}.

\bibitem{dershowitz82tcs}
N.~Dershowitz.
\newblock Orderings for term rewriting systems.
\newblock {\em Theoretical Computer Science}, 17:279--301, 1982.

\bibitem{jouannaud91lics}
J.-P. Jouannaud and M.~Okada.
\newblock Executable higher-order algebraic specification languages.
\newblock In {\em Proc. of LICS'91}.

\bibitem{jouannaud97tcs}
J.-P. Jouannaud and M.~Okada.
\newblock {A}bstract {D}ata {T}ype {S}ystems.
\newblock {\em Theoretical Computer Science}, 173(2):349--391, 1997.

\bibitem{jouannaud99lics}
J.-P. Jouannaud and A.~Rubio.
\newblock The {H}igher-{O}rder {R}ecursive {P}ath {O}rdering.
\newblock In {\em Proc. of LICS'99}.

\bibitem{jouannaud96rta}
J.-P. Jouannaud and A.~Rubio.
\newblock A recursive path ordering for higher-order terms in eta-long
  beta-normal form.
\newblock In {\em Proc. of RTA'96\em, LNCS 1103}.

\bibitem{jouannaud01draft}
J.-P. Jouannaud and A.~Rubio.
\newblock Higher-order recursive path orderings {\em "\`a la carte"}, 2001.
\newblock Draft.

\bibitem{kamin80}
S.~Kamin and J.-J. L\'evy.
\newblock Two generalizations of the {R}ecursive {P}ath {O}rdering, 1980.
\newblock Unpublished.

\bibitem{klop93tcs}
J.~W. Klop, V.~van Oostrom, and F.~van Raamsdonk.
\newblock Combinatory reduction systems: introduction and survey.
\newblock {\em Theoretical Computer Science}, 121:279--308, 1993.

\bibitem{kruskal60ams}
J.~B. Kruskal.
\newblock Well-quasi-ordering, the tree theorem, and vazsonyi's conjecture.
\newblock {\em Transactions of the American Mathematical Society}, 95:210--225,
  1960.

\bibitem{loria92ctrs}
C.~Loria-Saenz and J.~Steinbach.
\newblock Termination of combined (rewrite and $\lambda$-calculus) systems.
\newblock In {\em Proc. of CTRS'92\em, LNCS 656}.

\bibitem{lysne95rta}
O.~Lysne and J.~Piris.
\newblock A termination ordering for higher order rewrite systems.
\newblock In {\em Proc. of RTA'95\em, LNCS 914}.

\bibitem{mayr98tcs}
R.~Mayr and T.~Nipkow.
\newblock Higher-order rewrite systems and their confluence.
\newblock {\em Theoretical Computer Science}, 192(2):3--29, 1998.

\bibitem{miller89elp}
D.~Miller.
\newblock A logic programming language with lambda-abstraction, function
  variables, and simple unification.
\newblock In {\em Proc. of ELP'89\em, LNCS 475}.

\bibitem{plaisted78tr}
D.~A. Plaisted.
\newblock A recursively defined ordering for proving termination of term
  rewriting systems.
\newblock Technical report, University of Illinois, Urbana-Champaign, United
  States, 1978.

\bibitem{tait72lc}
W.~W. Tait.
\newblock A realizability interpretation of the theory of species.
\newblock In R.~Parikh, editor, {\em Proceedings of the 1972 Logic Colloquium},
  volume 453 of {\em Lecture Notes in Mathematics}, 1975.

\bibitem{vandepol93hoa}
J.~van~de Pol.
\newblock Termination proofs for higher-order rewrite systems.
\newblock In {\em Proc. of HOA'93\em, LNCS 816}.

\bibitem{oostrom93hoa}
V.~van Oostrom and F.~van Raamsdonk.
\newblock Comparing {C}ombinatory {R}eduction {S}ystems and {H}igher-order
  {R}ewrite {S}ystems.
\newblock In {\em Proc. of HOA'93\em, LNCS 816}.

\bibitem{walukiewicz03jfp}
D.~Walukiewicz-Chrz\k{a}szcz.
\newblock Termination of rewriting in the {C}alculus of {C}onstructions.
\newblock {\em Journal of Functional Programming}, 13(2):339--414, 2003.

\bibitem{walukiewicz03thesis}
D.~Walukiewicz-Chrz\k{a}szcz.
\newblock {\em Termination of Rewriting in the {C}alculus of {C}onstructions}.
\newblock PhD thesis, Warsaw University, Poland and Universit\'e d'Orsay,
  France, 2003.

\end{thebibliography}

\newpage
\tableofcontents

\end{document}